\documentclass[fleqn,10pt]{wlscirep}
\usepackage{color,graphicx}
\usepackage{dcolumn}
\usepackage{bm}
\usepackage{amsmath}
\usepackage{amsfonts}
\usepackage{amssymb}
\usepackage{textcomp}
\usepackage[position=top,caption=false]{subfig}

\begin{document}

\title{Correlation between topological band character and chemical
  bonding in a $\mathbf{Bi_{14}Rh_{3}I_{9}}$-based family of insulators}

\author[1,*]{Bertold Rasche}
\author[1]{Anna Isaeva}
\author[1,2]{Michael Ruck}
 \affil[1]{Department of Chemistry and Food Chemistry, TU Dresden, D-01062 Dresden, Germany} 
 \affil[2]{Max Planck Institute for Chemical Physics of Solids, D-01187 Dresden, Germany}

\author[3]{Klaus Koepernik}
\author[3]{Manuel Richter}
\author[3,+]{Jeroen van den Brink}
\affil[3]{Leibniz Institute for Solid State and Materials Research, IFW Dresden, D-01069 Dresden, Germany}

\date{\today}

\begin{abstract}
Recently the presence of topologically protected edge-states in Bi$_{14}$Rh$_3$I$_9$ was confirmed by scanning tunnelling microscopy consolidating this compound as a weak 3D topological insulator (TI). Here, we present a density-functional-theory-based study on a family of TIs derived from the Bi$_{14}$Rh$_3$I$_9$ parent structure via substitution of Ru, Pd, Os, Ir and Pt for Rh. Comparative analysis of the band-structures throughout the entire series is done by means of a unified minimalistic tight-binding model that evinces strong similarity between the quantum-spin-Hall (QSH) layer in Bi$_{14}$Rh$_3$I$_9$ and graphene in terms of $p_z$-molecular orbitals. Topologically non-trivial energy gaps are found for the Ir-, Rh-, Pt- and Pd-based systems, whereas the Os- and Ru-systems remain trivial. Furthermore, the energy position of the metal $d$-band centre is identified as the parameter which governs the evolution of the topological character of the band structure through the whole family of TIs. The $d$-band position is shown to correlate with the chemical bonding within the QSH layers, thus revealing how the chemical nature of the constituents affects the topological band character.
\end{abstract}

\maketitle
\section*{Introduction}
A promising feature of topological insulators (TIs)~\cite{kane_z_2_2005, kane_quantum_2005} is that their edge-states possess a helical locking of spin and momentum~\cite{qi_topological_2011, hasan_colloquium:_2010}. Not only does this feature strongly suppress back-scattering, it also lays the foundation for novel types of information processing such as spintronics~\cite{moore_birth_2010} by providing pure spin currents~\cite{brune_spin_2012}, or fault tolerant quantum computation by using the Majorana fermions at interfaces of topological states with superconductors~\cite{alicea_new_2012}. 
Manipulating such helical electrons becomes much easier if they
self-organise into quasi one-dimensional channels. Precisely this
occurs in topological insulators that consist of stacked quantum spin
Hall (QSH) layers~\cite{fu_topological_2007}, the rather ineptly named
weak TIs (WTIs).
In a WTI certain surfaces do not carry topological surface states. 
But it has been shown theoretically that these surfaces accommodate topologically protected spin-momentum locked edge-states at their step-edges~\cite{mong_quantum_2012, fu_topology_2012, morimoto_stability_2014}. Being confined to a step-edge, these chiral edge-states are intrinsically one-dimensional and can in principle be manipulated by changing the step-edge geometry~\cite{cano-cortes_fundamental_2013}. Very recently these electron channels have indeed been observed experimentally by scanning tunnelling microscopy~\cite{pauly_subnanometre-wide_2015} at the surface step-edges of Bi$_{14}$Rh$_3$I$_9$, the first experimentally realised weak 3D TI~\cite{rasche_stacked_2013}. 

\begin{figure}
\centering
\subfloat[Triclinic $\text{Bi}_{14}\text{Rh}_{3}\text{I}_{9}$]{\includegraphics[width=.35\columnwidth]{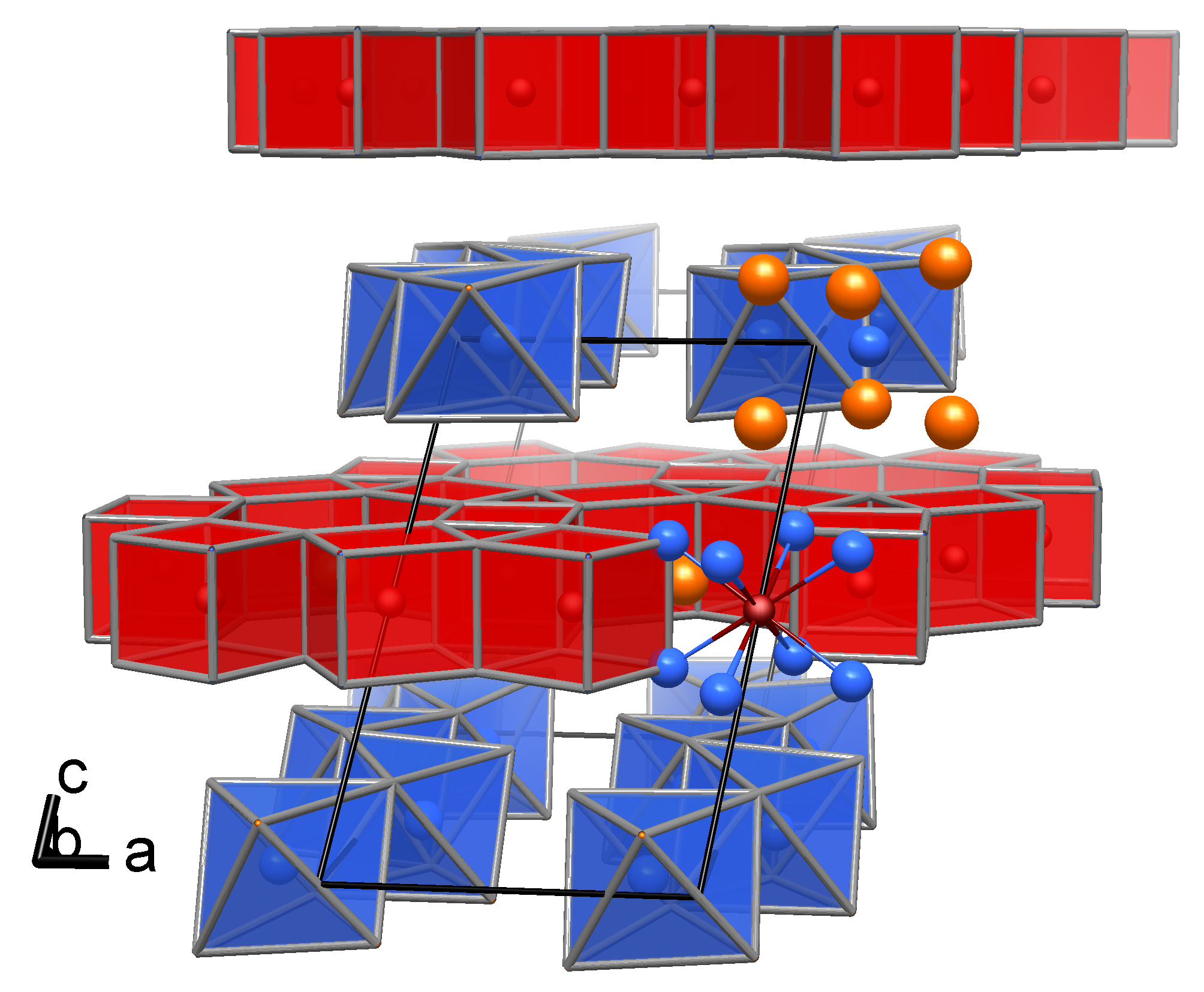}}
\subfloat[Hexagonal ILS]{\includegraphics[width=.35\columnwidth]{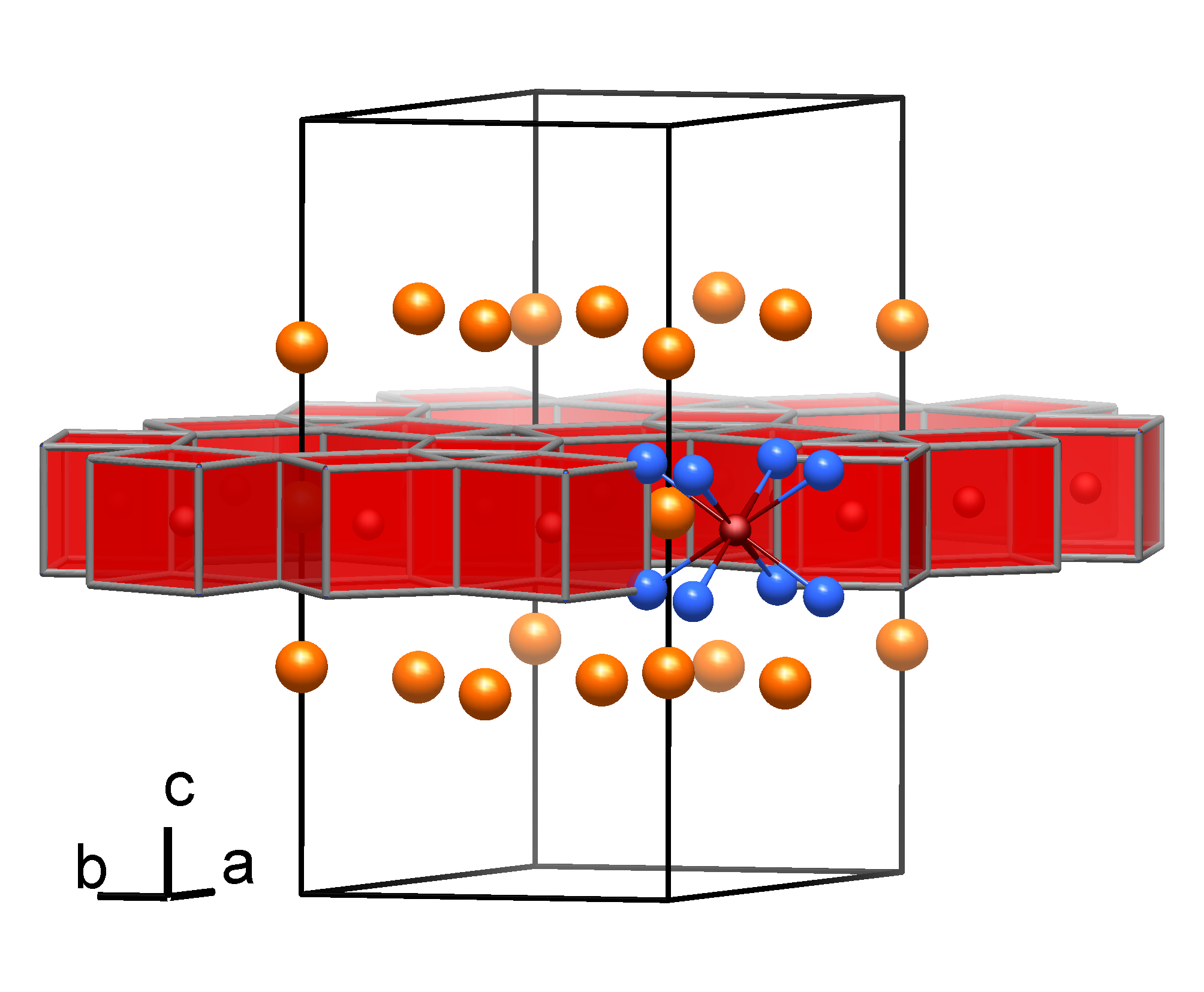}}
 \caption{\label{fig:structure} Structure of (a) triclinic
  $\text{Bi}_{14}\text{Rh}_{3}\text{I}_{9}$ (space group
  $P\overline{1}$) and (b) the corresponding hexagonal isolated-layer
  structure (ILS) (layer group $p6/mmm$). The
  QSH-layer, an intermetallic network of rhodium-centred
  bismuth-cubes, is coloured in red. Atom colours:   Bi: blue; I: orange; Rh: red.
}
\end{figure}

This layered compound features two structural fragments
(Figure~\ref{fig:structure}(a)). One is the QSH layer, a covalently bonded
intermetallic arrangement of edge-sharing, rhodium-centred
bismuth-cubes that resemble the honeycomb structure of graphene (Figure~\ref{fig:graphene-ILS}). 
Its hexagonal prismatic voids 
are occupied by iodide ions,
resulting in the layer composition [(Bi$_4$Rh)$_3$I]$^{2+}$. The
other layer, a spacer separating these 2D-TIs, has mainly ionic
character. There, bismuth-centred iodine octahedra share edges to form
[Bi$_2$I$_8$]$^{2-}$ chains. This experimentally realised stack is immune to translational disorder along the stacking direction
thanks to the high rigidity of the covalently-bonded QSH layer and the
electrostatic interaction of the oppositely charged QSH and spacer layers~\cite{rasche_crystal_2013}.

Identification of new TI-compounds is an important task for the
progress of the field. Since the TI-state is determined by the
character of the Bloch-states, chemical guidelines
towards synthesis of new TIs are vague hitherto. Yet, the innate structure-property interrelation should exist since the
Bloch-states, and therefore the bulk band-structure, reflect the bonding interactions of each given system. The QSH layer as found in $\text{Bi}_{14}\text{Rh}_{3}\text{I}_{9}$ is an ideal test system to address this issue. On the one hand, essentially the same intermetallic layer also exists with platinum instead of rhodium in the experimentally synthesised compounds $\text{Bi}_{13}\text{Pt}_{3}\text{I}_{7}$ and $\text{Bi}_{12}\text{Pt}_{3}\text{I}_{5}$~\cite{ruck_bi13pt3i7:_1997,kaiser_low-temperature_2014}, where the difference in electron count on the transition metal is counter-balanced by a different spacer. On the other hand, varying the interlacing spacer also changes the interactions between the intermetallic layers and affects the TI-properties~\cite{pauly_subnanometre-wide_2015}. This not only justifies substitution in the intermetallic layer, but also gives a broader perspective for physical experiments. From the viewpoint of synthesis, another advantage is that this class of materials is a fruitful playground for a versatile chemistry that stabilises low-dimensional networks in three-dimensional crystals, as different types of interactions coexist in them~\cite{ruck_metal_2001}. 
This paper addresses the questions whether $\text{Bi}_{14}\text{Rh}_{3}\text{I}_{9}$ can foster a whole family of TIs upon substitutions of transition-metal atoms and what the role of chemical interaction is, in the formation of the topologically non-trivial state. 

\section*{Results}
\textbf{Structural Modelling --} 
The overall structural complexity can be reduced by recovering the hexagonal symmetry of the QSH-layer [(Bi$_4$Rh)$_3$I]$^{2+}$. In the structure model, a single QSH-layer is isolated together with the iodine atoms of the [Bi$_2$I$_8$]$^{2-}$-spacer compatible with the QSH-layer symmetry $p$6/$mmm$. These iodine atoms are kept to compensate the charge. They are treated by the virtual crystal approximation (VCA) to guarantee the number of valence electrons to be as in the original compound.
The resultant model is referred to as isolated-layer
structure (ILS) or in the specific case with rhodium as Rh-ILS, throughout the text and is shown in Figure~\ref{fig:structure}(b). A density functional theory (DFT) based
calculation using the FPLO program
package~\cite{koepernik_full-potential_1999} shows that the main
signatures of the $\text{Bi}_{14}\text{Rh}_{3}\text{I}_{9}$ band-structure remain intact and that the
Rh-ILS is indeed a 2D-TI, confirming the adequacy
of the model (Details on the structural modelling and an extended comparison see Supporting Information).

\begin{figure*}
\begin{tabular*}{2\columnwidth}{ccc}
Ru & Rh & Pd \\
\includegraphics*[scale=0.24]{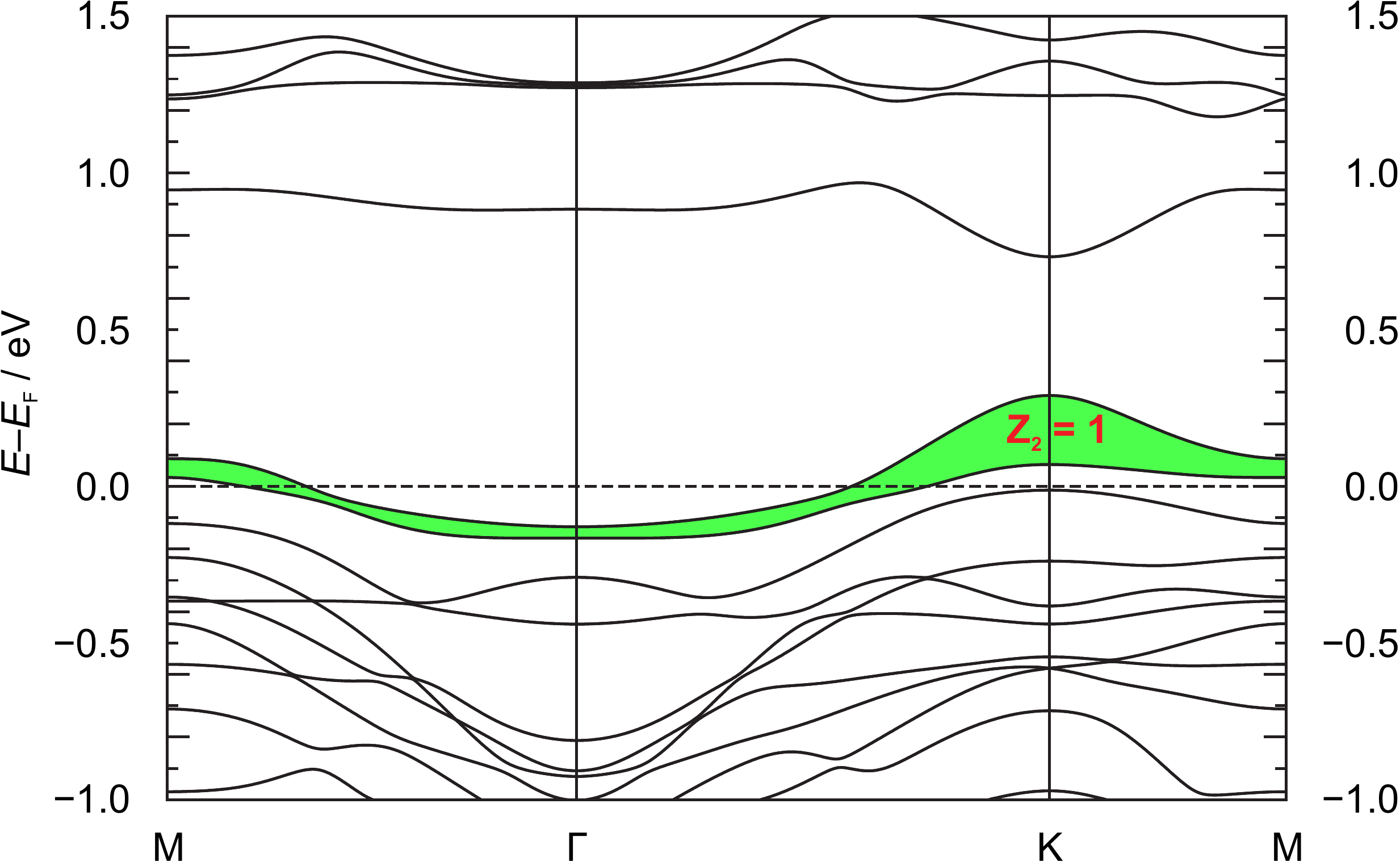}&
\includegraphics*[scale=0.24]{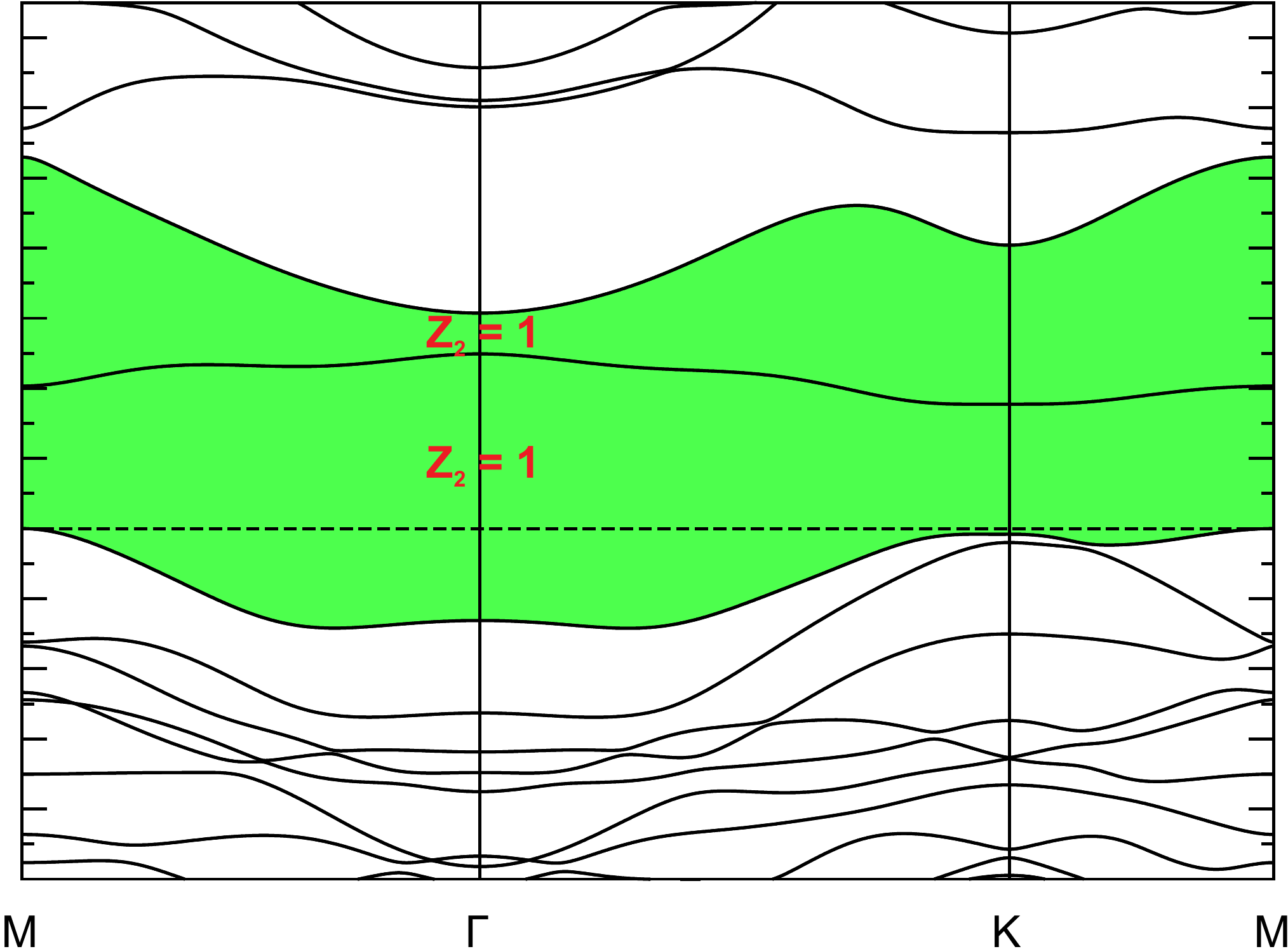}&
\includegraphics*[scale=0.24]{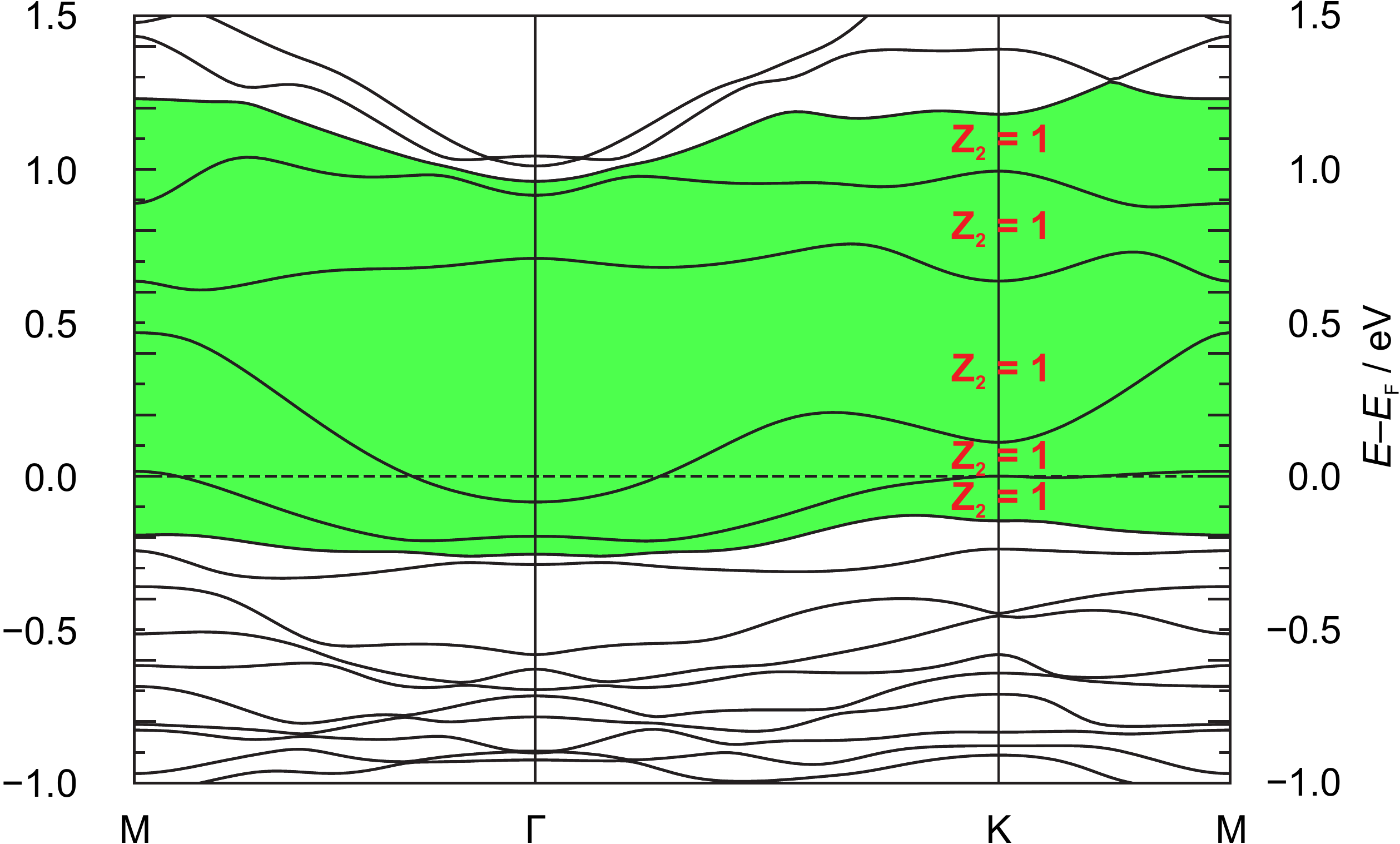}\\
Os & Ir & Pt \\
\includegraphics*[scale=0.24]{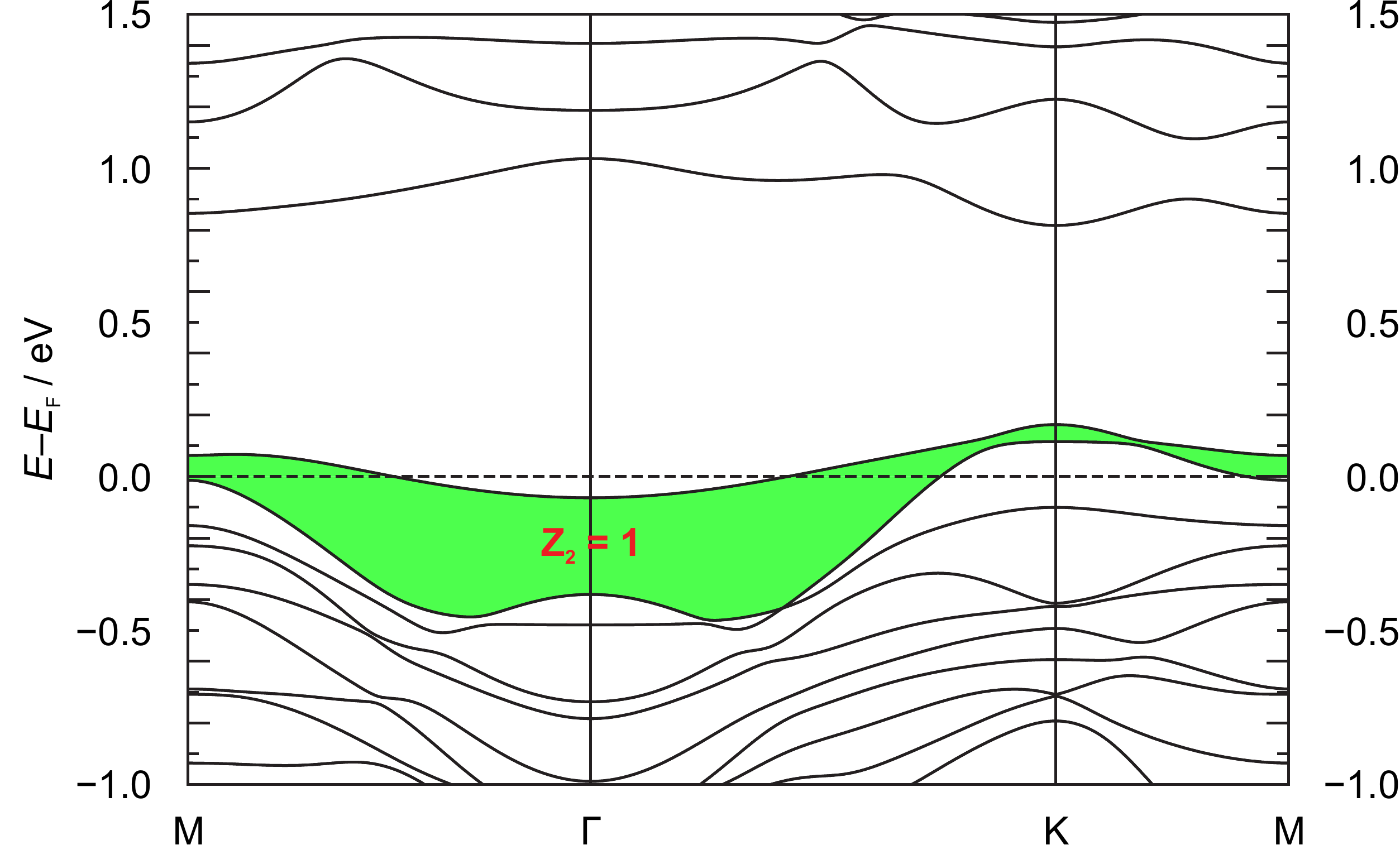}&
\includegraphics*[scale=0.24]{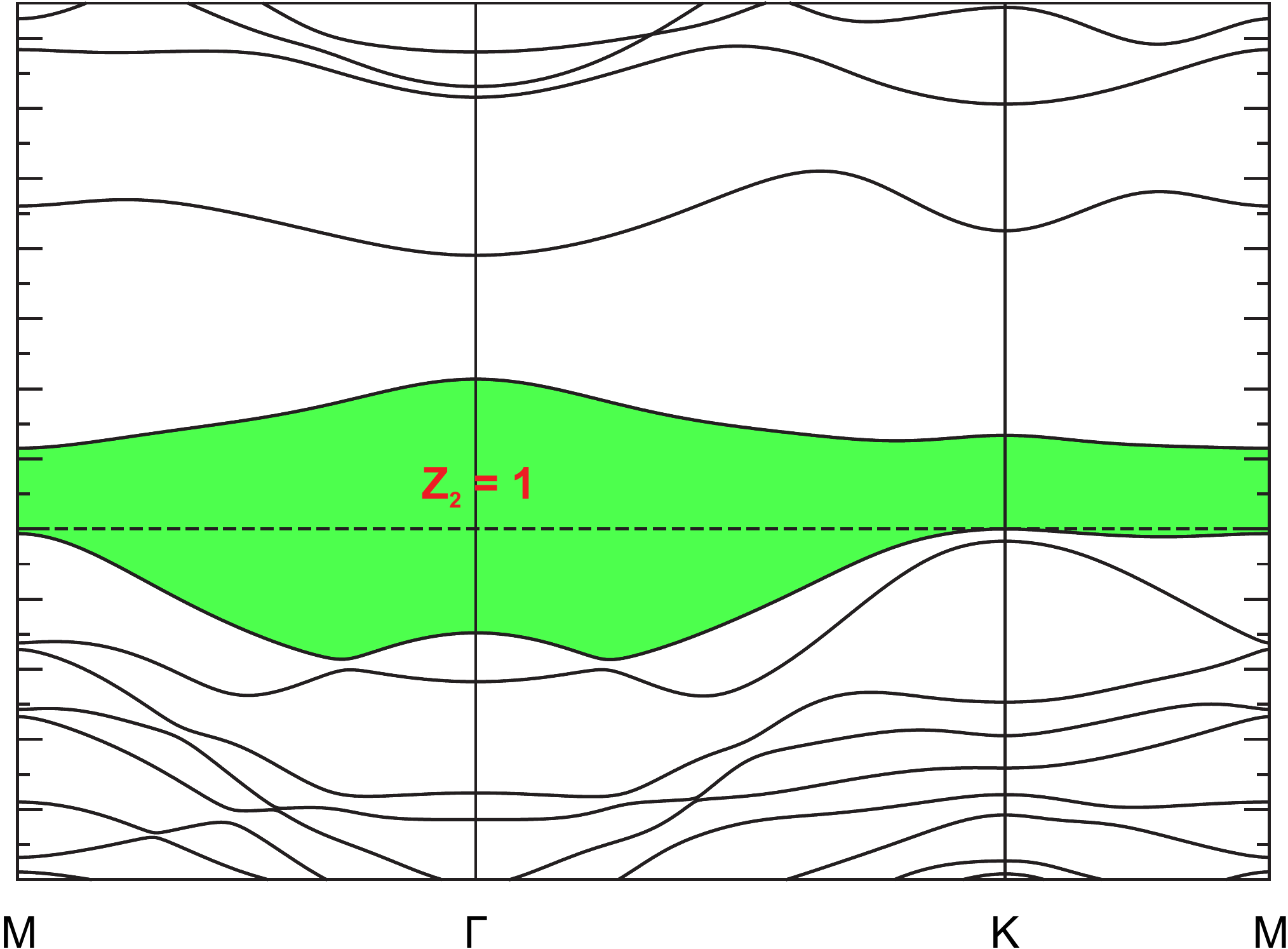}&
\includegraphics*[scale=0.24]{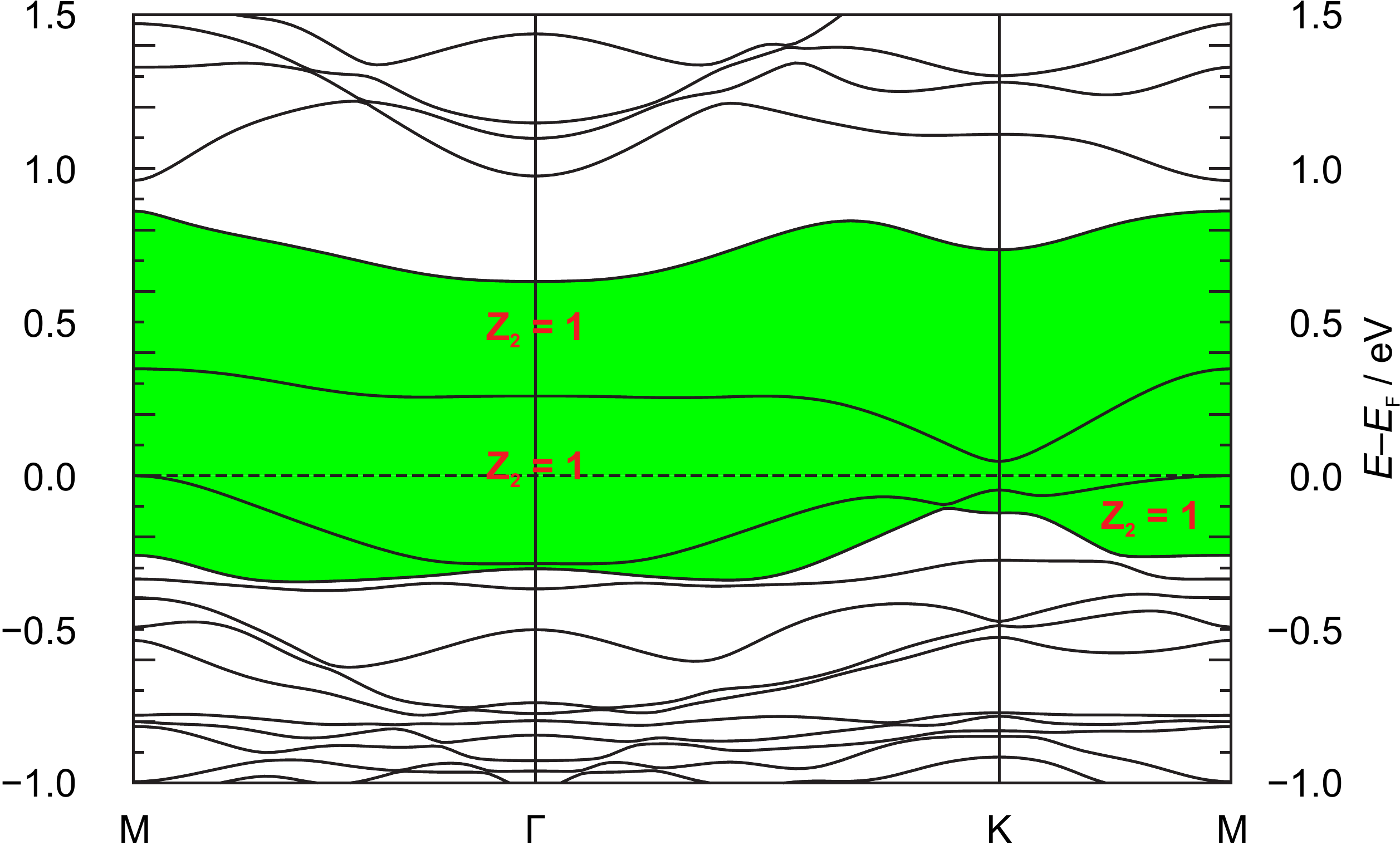}
\end{tabular*}
\caption{\label{fig:band_all}Full-relativistic band-structures for the
  ILS-series, all with a valence-electron count equal to the one in
  the original QSH-layer [(Bi$_4$Rh)$_3$I]$^{2+}$, arranged as in the
  periodic table of elements.
  The green shading depicts topologically non-trivial regions close to
  the Fermi-energy, $E_F$ (see explanation in the text).}
\end{figure*}

Metal substitutions were simulated by structural relaxation of M-ILSs
with M = Ru, Pd, Os, Ir, Pt. Only slight changes in the
interatomic distances were observed (SI: Figure~S3).
To a first approximation, the number of valence electrons in the QSH layer
was kept the same as in the original Rh-ILS by VCA applied to the iodine sites. For Ru and Os this implies a less, for Pd and Pt a higher charged M-ILS as compared to Rh and Ir, since the number of valence electrons increases with group number. Assumed charge flexibility of the M-ILS is plausible, since compounds with variously charged spacers and transition metals from different groups have already been synthesised. Nevertheless, other possible electron counts for the M-ILSs were computationally addressed as well and are discussed below. Scalar- and full-relativistic band-structures and the topological invariants were subsequently calculated
for all M-ILSs (Figure \ref{fig:band_all} and SI: Figure~S4, Figure~S6 and Table~S1).

\textbf{Electronic Structure --} 
For full-relativistic band-structures, a topologically non-trivial
``warped'' gap of varying width can be found for all M-ILSs
(Figure \ref{fig:band_all}).
The term ``warped'' refers to the fact that two bands are non-crossing and therefore have an 
energy separation at each $k$-point in the Brillouin zone. Yet, as they can have an energy overlap at \textit{different} $k$-points this ``warped'' gap is not necessarily an actual energy gap, even if the ``warped'' band structure is adiabatically connected to a fully gapped electronic state.

The bands that change the $Z_2$-invariant from trivial to non-trivial
or vice versa we call ``topologically active'' bands, whereas the ones
that do not change $Z_2$ are called ``topologically in-active''
bands.
Formally topological activity can be calculated separately for all (even crossing) bands.
The change in topological nature, however, only manifests itself in a warped or real gap above
the last active band (phase continuity condition for non-crossing band complexes).

As shown in Figure \ref{fig:band_all}, only in case of the Rh-,
Ir-,Pt- and Pd-ILS an actual band gap of the material falls into the
discussed topologically non-trivial region in the vicinity of the
Fermi-energy. Hence these ILSs should support the 2D-TI layers given a
certain number of valence-electrons. On the contrary, the Ru- and
Os-ILS have only a large trivial gap, while the topologically active
bands encompass a non-trivial warped gap.

The Rh-, Ir- and Pt-ILS are found to be insulating, with a valence-electron count as found in the original $\text{Bi}_{14}\text{Rh}_{3}\text{I}_{9}$ structure. On the contrary, in the cases of M = Ru, Os
and Pd addition of
two electrons per formula unit is necessary to achieve insulating
behaviour. We established via additional calculations that the ILS
band-structure changes only insignificantly for such an amount of electron
doping (SI: Figure~S6). The same number of electrons added to the
Rh-, Ir- and Pt-ILS will drive these systems from one insulating state
to another, the latter being topologically non-trivial for the Rh- and
Pt-cases.

The described differences in the electronic band-structures obviously
arise from the $d$-states of the transition elements: these are the
valence states that vary from one M-ILS to the other.
Generally, the energy position of the $d$-band within a period becomes lower
with increasing group number of the M-element.
Thus, in the ILS-series the energies of the populated $d$-states are
also lowered with increasing group number, which means that the
contribution of the $d$-states to the density of states (DOS) close to
the Fermi-energy progressively decreases (SI: Figure~S5).

\textbf{Origin of the topological nature --} 
In order to uncover the mechanism that drives the ILS systems into a
topologically insulating state, 
we extracted the appropriate tight-binding (TB) models from the calculated band structures and analysed the influence of M-element substitutions on the topological band character. 
The TB-description includes the Bi
$p_z$- and $p_{xy}$-states, and the M $d$-states (details in SI). Considered are nearest neighbour
interactions for each pair of elements.
Switching all of them on step-by-step, we arrived at a
well-reproduced scalar-relativistic band-structure for the Rh- or any other ILS
(SI: Figure~S9).

The sets of model parameters extracted from scalar-relativistic
FPLO-based calculations were found rather similar for all ILS. The
decisive differences arise from the position of the onsite
energies of the M $d$-orbitals relative to the onsite energies of the Bi $p_z$ and $p_{xy}$-orbitals. Moreover, the similarity of all band-structures enables
us to derive all other models from the Rh-ILS by simply shifting the
onsite energies of the M $d$-states with respect to Bi $p$-orbitals
(SI: Figure~S13). As a result, the evolution of
the band-structure upon M-substitutions can be traced qualitatively and quantitatively by
means of just one parameter, the $d$-state onsite energy: 
the lowering of the $d$-state energies is chiefly responsible for the changes
in the ILS-band-structure. Ordering from the highest
onsite $d$-energy to the lowest yields the sequence Os, Ru, Ir, Rh, Pt
and Pd.

\begin{figure}
\centering
\includegraphics[width=0.5\columnwidth]{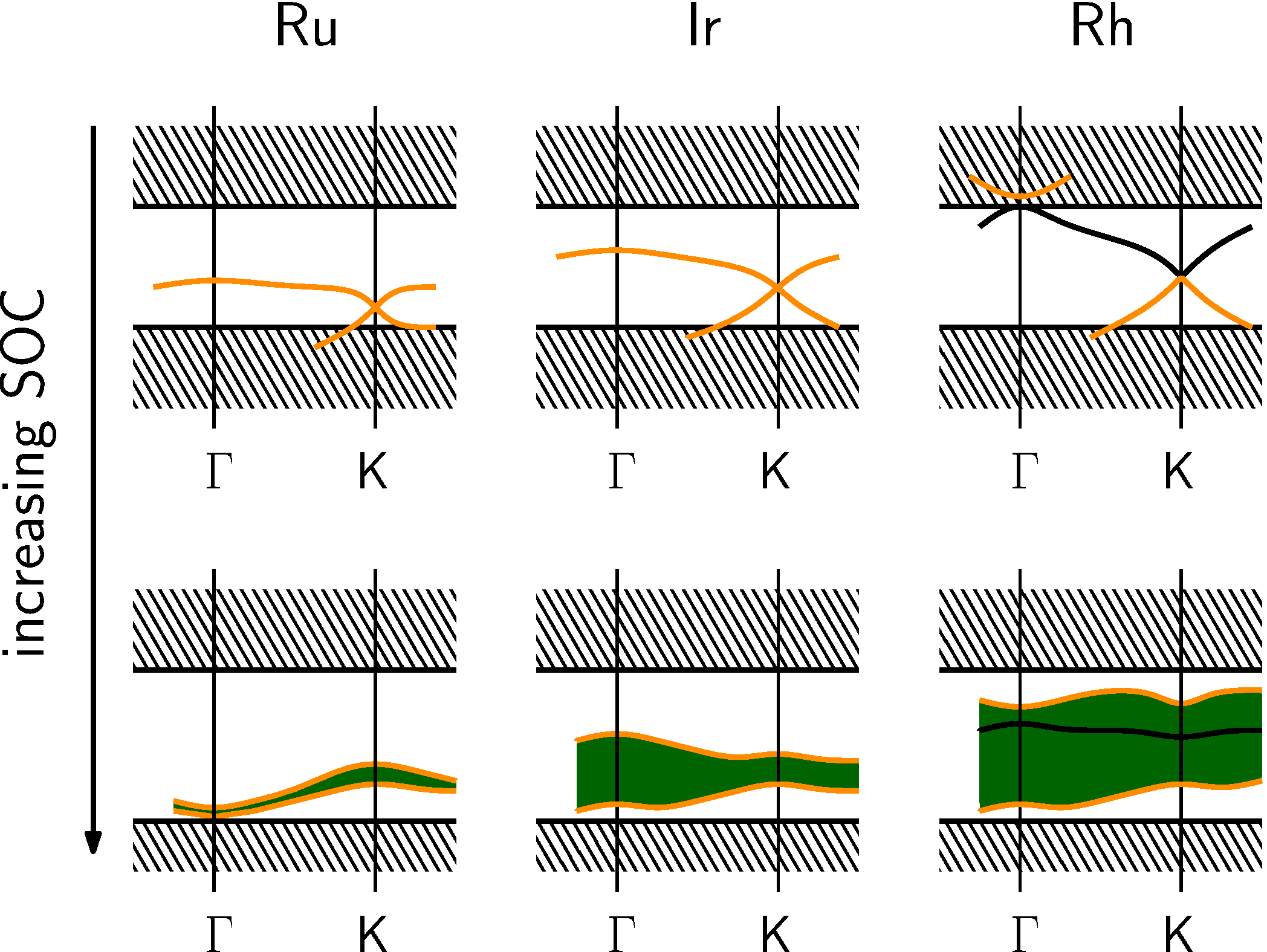}
\caption{\label{fig:topogap} Sketch of the evolution of the scalar- into the full-relativistic band-structure by switching on the spin-orbit-coupling. The shaded areas represent the valence and conduction band continua. Orange bands are topologically active. Black bands are in-active (explanation see text). The green area is topologically non-trivial.}
\end{figure}

Subsequently, the emergence of the topologically non-trivial state was
analysed by switching on spin-orbit-coupling (SOC).
Continuously increasing the strength of the SOC,
we traced the evolution of the scalar-relativistic bands into their
full-relativistic counterparts. Figure \ref{fig:topogap} presents a
scheme of the most discernible situations.

For Os and Ru the topologically active bands that form a cone at the K
point in the scalar-relativistic band-structure reside 
near the bottom of a large gap. When the SOC is switched on, these crossing bands gap out, thus
forming a very narrow topological region warped in the
$k$-space. However this region is located below a larger trivial gap that
determines the trivial insulating character of the Ru- and Os-based
model compounds (if charged with two additional electrons).
For Ir, the dispersion of the cone-forming bands is
larger, so that a sizeable topologically non-trivial gap appears with
SOC.

The remaining Rh-, Pt- and Pd-ILS fall into an essentially different
category. Due to the wide energetic spread of the cone-forming bands a
second band crossing occurs at the $\Gamma$-point. This additional
crossing affects the topological activity of the adjacent bands
and the upper part of the cone at the $\Gamma$ point becomes topologically
active. The SOC gaps out both cones so that the topologically
non-trivial region becomes even broader in energy than in the Ir-ILS and topologically
in-active bands reside within this region.

From Rh to Pd more and more in-active bands are found within the topologically non-trivial region. Most likely the increasing  energetic difference of the Bi $p_z$-$p_z$ anti-bonding and bonding states allows for this trend, which can be traced by the spectral weight in the TB-model. 

\begin{figure}[tb]
\centering
\subfloat[\label{fig:elid}]{\includegraphics[width=.35\columnwidth]{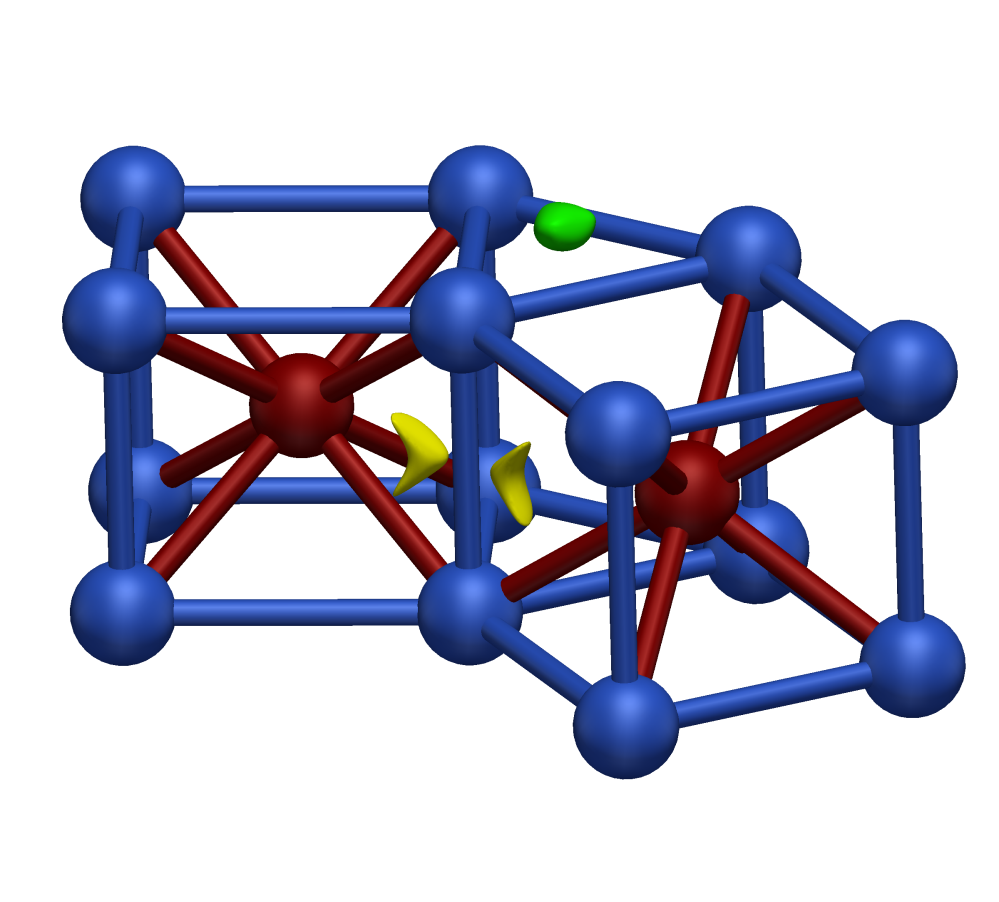}}\hspace{1cm}
\subfloat[\label{fig:popu}]{\includegraphics[width=.35\columnwidth]{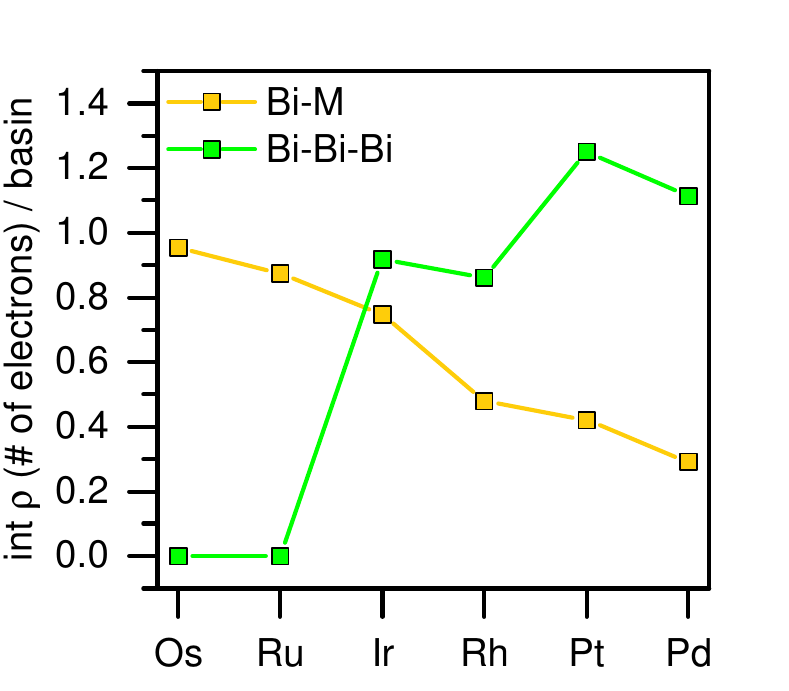}}
\\
\subfloat[\label{fig:graphene-ILS}]{\includegraphics[width=0.35\columnwidth]{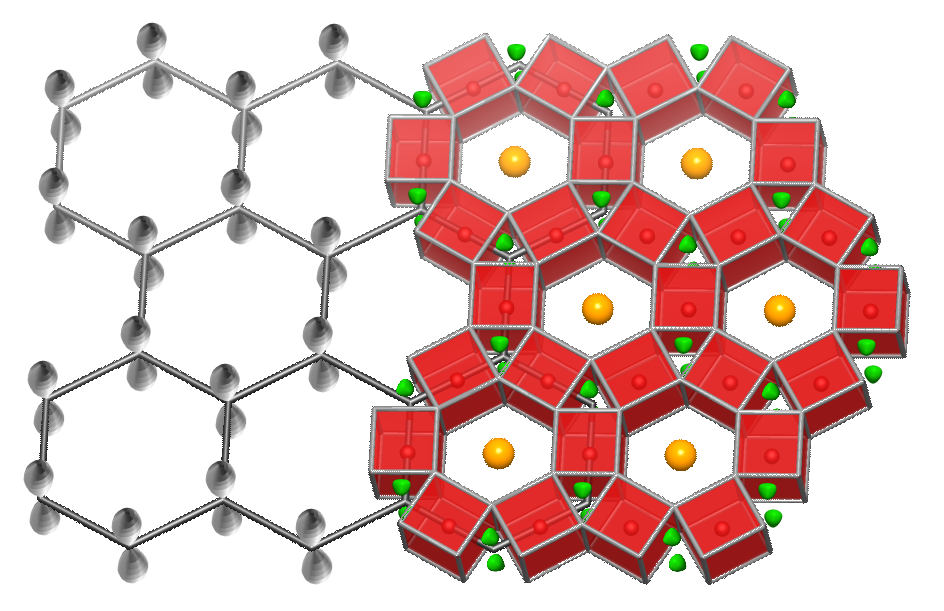}}\hspace{1cm}
\subfloat[\label{fig:G-point-wave-function}]{\includegraphics[width=0.35\columnwidth]{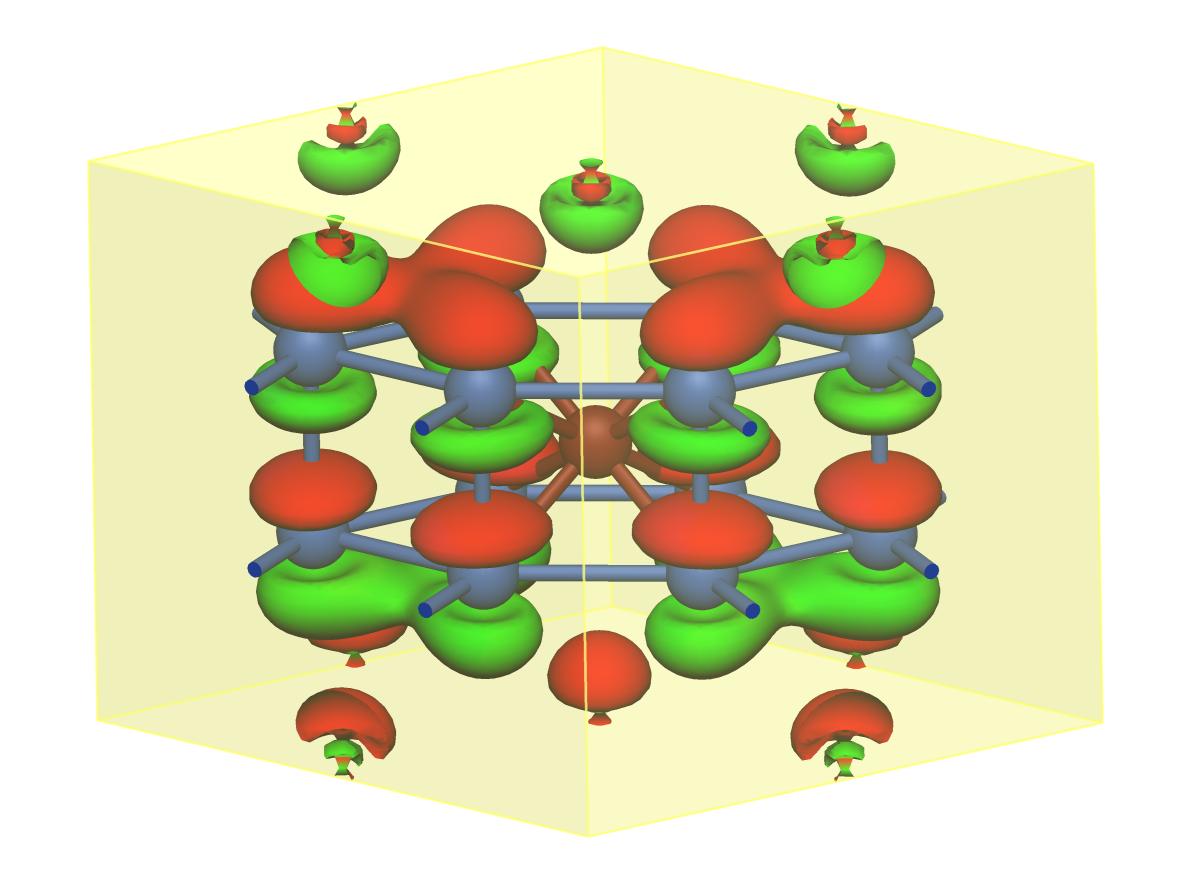}}
\caption{
(a) Two selected ELI-D localisation domains shown within an ILS structural fragment. One domain is depicted for each basin type: the Bi--M-bond (yellow) and the three-centre bismuth bond (green).\\
(b) Integrated electron density within these ELI-D basins for all M-ILSs (same colour code as for (a)).\\
(c) Overlay of structure of graphene (left, scaled by a factor of approx. 3.8) with sketches of the $p_z$-Orbitals (grey) and the ILS (right) with the ELI-D localisation domains of the three-centre bismuth bond (green isosurface). \\
(d) Structural fragment with two Bi$_6$ prisms, indicating the spin-up component of the $\Gamma$-point wave function of a band with anti-bonding triangular MO weight. On each prism six Bi $p_z$ orbitals combine into a single pseudo $p_z$ orbital.
}
\end{figure}

Complementary, this trend of decreasing onsite
energies of the $d$-states is also reflected in real space in the chemical bonding,
which was visualised and quantified via analysis of the electron localisability indicator
(ELI-D) field. Previously this approach revealed
two types of bonding basins for the triclinic structure of
$\text{Bi}_{14}\text{Rh}_{3}\text{I}_{9}$. One was assigned to Rh--Bi
interactions and the other corresponded to a three-centre bismuth
interaction~\cite{rasche_stacked_2013}. The first type of ELI-D basin
is present in all six M-ILSs (yellow isosurface in
Figure~\ref{fig:elid}) and accounts for the rigidity of this
single intermetallic layer. Basin population, i.e.\ the number of
electrons per bond, decreases in the same sequence in which the
onsite energies lower from Os to Pd
(Figure~\ref{fig:popu}). Ru and Os interact most strongly with bismuth, while 
Pt and Pd, which have their $d$-states filled, draw less electrons
from the bismuth atoms and show weaker interactions.

\textbf{Graphene with strong spin-orbit-coupling --}
The second type of ELI-D bonding feature
is {\it only} present in the ILS with Rh, Ir, Pd, and Pt, i.e. in the
topologically non-trivial phases. The basin attractor is located
slightly off the plane formed by three bismuth atoms in both bases of each bismuth
triangular prism (see green isosurface in
Figure~\ref{fig:elid}). The evaluation of basin population
shows that the strength of this bonding increases within the group and with
the period number. The three-centre Bi--Bi bonding intensifies with
the weakening of the M--Bi bonding, leaving more electrons on the
Bi-site for homoatomic interaction. Thus, a correlation with the population of the M
$d$-states can be established: Ru and Os form strong
M--Bi bonds and contribute significantly to the states near the
Fermi-level, while Pd and Pt retreat from the Fermi
level, which is then dominated by Bi--Bi bonding.

The intriguing finding is that at least one topologically non-trivial
energy gap is found for all ILSs that show this three-centre bismuth
bonding. The importance of this three-centre bismuth bond becomes even
more striking when comparison is made between the hexagonal
arrangement in the ILSs and graphene (Figure \ref{fig:graphene-ILS}). The position of the basin attractors
matches with the positions of the carbon atoms. 
The topologically
non-trivial M-ILSs can thus be considered as graphene analogues in a
sense that three-centre bismuth bonds reside in place of the carbon
atoms. As these bonding attractors are found on both sides of the QSH
layer one 
might conclude that these basins can be compared to
the lobes of the $p_z$-orbitals of the carbon atoms in
graphene. Having a minimal TB-model at hand, we can evaluate whether
these real-space similarities of the M-ILSs and graphene are
also reflected in the electronic structure. The topological properties
of the graphene lattice arise from the Dirac-cone formed by
$p_z$-orbitals, upon inclusion of SOC. 
The topological nature of the M-ILSs should similarly be related
to a lattice arrangement of molecular orbitals (MOs) which are contributed
by several bismuth atomic orbitals. An analysis of the
full-relativistic band-structure reveals that the Dirac-like cone
occurring at the K-point is indeed related to an effective MO which combines six of the Bi $p_z$-orbitals into a
single pseudo-$p_z$-orbital (Figure~\ref{fig:G-point-wave-function}). Therefore the cone observed at the Fermi-level in the
ILSs is to a notable extent analogous to the Dirac-cone in graphene.
Nevertheless it should be emphasised that
the evolution of the topological gaps in the series
cannot be solely attributed to Bi-$p_z$ spectral weight.
The situation in the ILSs is more complex: although the Dirac-cone in the scalar-relativistic
band-structure has significant Bi $p_z$-character, 
hybridisation with other states takes place, giving rise to further topologically active bands. 

\section*{Discussion}
A simplified structural approximant -- an isolated-layer structure --
was constructed for the first experimentally realised weak 3D TI 
$\text{Bi}_{14}\text{Rh}_{3}\text{I}_{9}$. 
This model retains all essentials of the initial electronic structure
and thus allows to explore a family of TI-candidate materials obtained 
by a systematic replacement of Rh by M = Ru, Os, Ir, Pd, Pt.
Besides the Rh-ILS, the systems with Pd, Ir, or Pt exhibit at least 
one topologically non-trivial gap near the Fermi-level for experimentally
achievable electron counts, whereas the Ru- and Os-substituted systems 
are topologically trivial.
The devised TB-model unravels the essential role of the transition metal 
$d$-band position for the topological character of the bands, as
this position
determines the composition and spread of the Bi derived bands at
and above the Fermi level.
The Bi states compose the topologically active bands that form cones at the $\Gamma$- and K-points in the
scalar-relativistic band-structure. SOC opens them
into a topologically non-trivial gap. The MO built by the Bi $p_z$-orbitals can be
regarded as a pseudo $p_z$-orbital which 
exemplifies 
the
resemblance between the QSH layer in
$\text{Bi}_{14}\text{Rh}_{3}\text{I}_{9}$ and graphene beyond purely
structural aspects. These signatures coincide with the occurrence of
three-centre bismuth bonds in case of the electron-rich transition metals, where
the bonding between bismuth and the M-atoms is weak.
Bismuth--bismuth interaction found in the TB-model as well as
from the real-space bonding analysis correlates with the topological
non-triviality of a model system and thus might be regarded as a
fingerprint of the TI state in this family of compounds.
It would be interesting to explore, whether a similar correlation
between the electronic structure/the bonding pattern and the presence
of a topological state can be found in other families
of (weak) topological insulators.

\section*{Methods}
Density functional theory (DFT) based calculations were performed with the full-potential local orbital (FPLO) code \cite{koepernik_full-potential_1999} version 14.00, within the local density approximation (LDA) \cite{perdew_accurate_1992}. For full-relativistic calculations the four-component Dirac equation was solved. The Bl\"ochl corrected linear tetrahedron method with a 12x12x1 k-point mesh for the ILSs and with a 10x6x6 k-point mesh for $\text{Bi}_{14}\text{Rh}_{3}\text{I}_{9}$ was employed for integration. The following basis states were treated as valence states: Bi: 5s, 5p, 5d, 6s, 6p, 6d, 7s,
7p; Ru: 4s, 4p, 5s, 6s, 4d, 5d, 5p; Rh: 4s, 4p, 5s, 6s, 4d, 5d, 5p; Pd: 4s, 4p, 5s, 6s, 4d, 5d, 5p; Os: 4f, 5s, 5p, 6s, 7s, 5d, 6d, 6p; Ir: 4f, 5s, 5p, 6s, 7s, 5d, 6d, 6p; Pt: 5s, 5p, 5d, 6s, 6p, 6d, 7s; I: 4s, 4p, 4d, 5s, 5p, 5d, 6s, 6p.

\subsubsection*{Unfolded band-structure}
For better comparison with the band-structures of the ILS, we unfold the band-structure of $\text{Bi}_{14}\text{Rh}_{3}\text{I}_{9}$. The unfolded band-structure was calculated following the procedure described in supplementary material of Ref.~\cite{van_heumen_existence_2011}. It is achieved by projecting the triclinic Bloch functions onto Bloch projectors of idealised hexagonal translational symmetry. The resulting band weights remove bands backfolded by the triclinic doubling of the real unit cell. This can only be done for the atoms of the intermetallic layer, since the spacer layer completely breaks the hexagonal symmetry.

\subsubsection*{Real-space bonding indicator, ELI-D}
Chemical bonding was characterised via the electron localisability indicator (ELI-D, $\Upsilon^\sigma_D$)\cite{kohout_measure_2004, kohout_bonding_2007}. The ELI-D 3D field was computed from the converged SCF calculations using the ELI-D module~\cite{ormeci_electron_2006} for the FPLO program package~\cite{koepernik_full-potential_1999}. The field maxima (attractors) can mark out various bonding features such as atomic shells, lone pairs and chemical bonds. Topological analysis of ELI-D was performed in the DGrid 4.6~\cite{kohout_dgrid_2009} program by a procedure similar to that of the quantum theory of atoms in molecules by Bader~\cite{bader_atoms_1990}. The integration of electron density within the resultant basins yielded the electron count for all bonding features. The atomic contributions into the bonding basins were computed via integration of the intersection between the bonding ELI-D basin and the respective QTAIM basins for atomic constituents. The ELI-D localisation domains were visualised in the real space in Paraview program package~\cite{_paraview:_2008}. Worth noting is the mutual consistency of the bonding patterns computed for triclinic $\text{Bi}_{14}\text{Rh}_{3}\text{I}_{9}$ using a full-potential full-electron code~\cite{rasche_stacked_2013} and for Rh-ILS using a full-potential local-orbital one (present work).

\section*{Acknowledgements}
We acknowledge the financial support from the German science foundation in the framework of the Special Priority Program {\it Topological Insulators} (SPP 1666) and the Collaborative Research Center {\it Correlated Magnetism: From Frustration To Topology} (SFB 1143). We are indebted to ZIH TU Dresden for the provided computational facilities. 

%

\end{document}